\documentclass[twocolumn,amsmath,prb,superscriptaddress]{revtex4-1}

\usepackage{graphicx}
\usepackage{dcolumn}
\usepackage{bm}
\usepackage{color}
\usepackage{subfigure}
\usepackage{hyperref}
\usepackage{url}
\usepackage{lineno}

\hypersetup{colorlinks=true, citecolor=blue,pdfborder={0 0 0}}

\makeatother

\begin{document}
\title{Cubic Dirac and quadruple Weyl points in screw-symmetric materials}

\author{Peng-Jen Chen}
\email{pjchen1015@gmail.com}
\affiliation{Institute of Physics, Academia Sinica, Taipei 11529, Taiwan}
\affiliation{Institute of Atomic and Molecular Sciences, Academia Sinica, Taipei 10617, Taiwan}
\author{Wan-Ju Li}
\email{wjli78@mail.nsysu.edu.tw}
\affiliation{Institute of Physics, Academia Sinica, Taipei 11529, Taiwan}
\affiliation{Department of Physics, National Sun Yat-sen University, Kaohsiung 80424, Taiwan}
\author{Ting-Kuo Lee}
\affiliation{Institute of Physics, Academia Sinica, Taipei 11529, Taiwan}
\affiliation{Department of Physics, National Sun Yat-sen University, Kaohsiung 80424, Taiwan}
\begin{abstract}
  High-order topological charge is of intensive interest in the field of topological matters. In real materials, cubic Dirac point is rare and the chiral charge of one Weyl point (WP) has never be found to exceed $|C|=3$ for spin-$\frac{1}{2}$ electronic systems. In this work, we argue that a cubic Dirac point can result in one quadruple WP ($|C| = 4$ with double band degeneracy) when time-reversal symmetry is broken, provided that this cubic Dirac point is away from the high-symmetry points and involves coupling of eight bands, rather than four bands that were thought to be sufficient to describe a Dirac point. The eight-band manifold can be realized in materials with screw symmetry. Near the zone boundary along the screw axis, the folded bands are coupled to their ``parent'' bands, resulting in doubling dimension of the Hilbert space. Indeed, in $\varepsilon$-TaN (space group 194 with screw symmetry) we find a quadruple WP when applying a Zeeman field along the screw axis. This quadruple WP away from high symmetry points is distinct from highly degenerate nodes at the high-symmetry points already reported. We further find that such a high chiral charge might be related to the parity mixing of bands with high degeneracy, which in turn alters the screw eigenvalues and the resulted chiral charge.
\end{abstract}

\maketitle

\section{Introduction}
Topological materials can be classified into topological insulators \cite{Zhan,Hasa,Qi}, topological crystalline insulators \cite{Fu1,Hsie,Tana,Okad,Ma}, topological nodal-line (TNL) systems \cite{Bian,Lian,Bia2}, Weyl semimetals \cite{Wan,Xu,Fang,Hua1,Weng,Lv,Xu2,Hua2,Deng,Jian,Wen2}, topological Dirac semimetals \cite{Wang,Liu,Liu2,Xion,Yan1,Liu3}, and so on. Each classification is characterized by its own topological invariant and may be featured by different behaviors of the bulk-edge correspondence. Among these, Dirac/Weyl semimetals have received growing interest in the field of topological materials. Many materials have been demonstrated to host Weyl points (WPs), such as TaAs (and its isostructural compounds) \cite{Hua1,Weng,Lv,Xu2}, $\mathrm{MoTe_2}$ \cite{Deng,Jian}, $\mathrm{HgCr_2Se_4}$ \cite{Xu,Fang}, $\mathrm{SrSi_2}$ \cite{Hua2} and so on. Most of the WPs in these known materials carry single chiral charge ($C = \pm 1$), except for $\mathrm{HgCr_2Se_4}$ and $\mathrm{SrSi_2}$ which are proposed to be a double-Weyl semimetal ($C = \pm 2$) \cite{Xu,Fang,Hua2}. The multi-Weyl semimetals are physically appealing for they are theoretically proposed to exhibit unusual physical properties depending on $|C|$ \cite{Dant,Ezaw,Gorb,Sun,Lu,Lu}. Cubic Dirac fermions are predicted to exist in quasi-one-dimensional transition-metal monochalcogenides \cite{Liu3}. Once inversion and/or time-reversal symmetry is broken, triple WPs with $C = \pm 3$ are expected to appear. It is then naturally to ask if we can find real materials with $|C|>3$? 

High degeneracy at high symmetry points has been proved to be a way to realize WPs with high chiral charges. There have been several works reporting the presence of "crossing points" with total chiral charge amounting to four \cite{bradlyn2016,schroter2019,chang2017,Wu}. These reported "crossing points" are highly degenerate
states on high-symmetry points protected by crystalline
symmetry. Clearly speaking, they are composed of two
WPs with total chiral charge being $2 + 2 = 4$ \cite{Wu} or $3 + 1 = 4$ \cite{bradlyn2016,schroter2019,chang2017}. Alternatively, Zhang et al. \cite{zhang2020} propose that the integer-spin systems can be a paradigm for preparing high-order WPs. They first discuss the doubly degenerate Weyl points at time-reversal-invariant momenta, such as $\Gamma$, with $|C|=4$ in the spinless electrons, photons, phonons and magnons without the spin-orbit couplings (SOC). After turning on the SOC, the doubly degenerate WP with $|C|=4$ disappears and a four-fold degenerate quadruple WP emerges. As the SOC is ubiquitous in electronic systems, it is then interesting to ask whether a spin-1/2 system with SOC can have doubly degenerate WP with a high chiral charge, such as $|C|=4$, in generic points (not high-symmetry points) in momentum space?


The current analyses of four-fold degenerate Dirac (doubly degenerate Weyl) point in SOC systems rely on the minimal four- (two-) band model \cite{Fang,Yan1}. Such a minimal model suffices because all states outside of this manifold are irrelevant. However, if more complicated coupling of bands exists, the dimension of the minimal model may need to increase. Moreover, once couplings between bands within a larger manifold occur, different arrangement of band crossings could produce different topological behaviors even though these bands obey the same space group symmetry. One possibility of generating a large manifold of bands is closeness to high-degenerate states. Therefore, we look for candidate systems in which high-degenerate states are nearby with or without SOC.

Recently, it is proposed that open straight TNLs with four-fold degeneracy would exist, on the $k_z = \pi/c$ plane, in materials belonging to space group 194 \cite{Lian,Yan2}. The non-symmorphic symmetry plays the essential role in protecting the bands along $A = (0, 0, \pi/c)$ to $L = (0, \pi/a, \pi/c)$ from spin-orbit splitting. The high-symmetry $A$ point is also where band folding (along $k_z$) occurs due to the screw symmetry. When two such folding points are present and close to each other in energy, there can exist a Dirac point on the screw axis away from $A$. Indeed, such Dirac point can exist in materials belonging to space group 194 when two such folding points are close in energy. In this work, we take $\varepsilon$-TaN for demonstration and find that such a Dirac point can be cubic. When a Zeeman field is applied along the screw axis, the Dirac point splits into four WPs with $C = \pm 1, \pm 2$. More interesting is that a WP with
$|C| = 4$ is formed nearby due to an additional band crossing. 
For comparison, we have also studied $\mathrm{Na_3Bi}$ that belongs to the same space group 194. Similar eight-band manifold is found in $\mathrm{Na_3Bi}$ from a set of bands lying $\sim$ 3 eV above the Fermi level. Having the same irreducible band representations, we find $|C| = 2$, instead of $4$, for the same band crossing in $\mathrm{Na_3Bi}$. This indicates that symmetry group and band representation are not sufficient to determine the chiral charge.

\section{Results}
\begin{figure}[h]
\includegraphics [width=7.8cm] {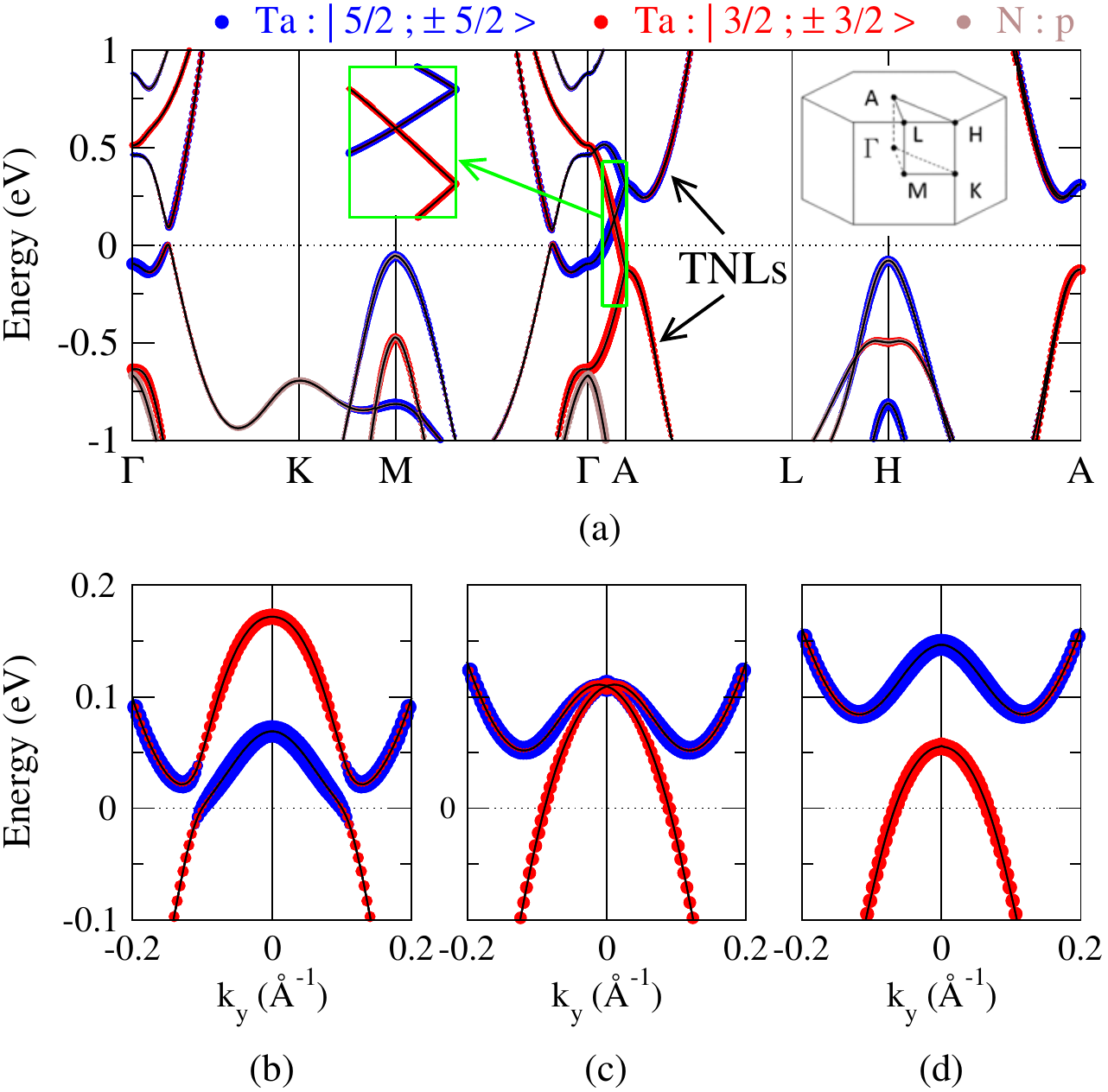}
\caption{\small{Orbital-decomposed band structure of $\varepsilon$-TaN. (a) The full band structure along the conventional high symmetry path as the right inset depicts. The left inset shows the Dirac point on the rotation axis. TNLs are marked by the arrows. (b)-(d) The band structure along $k_y$ direction around the Dirac point with (b) $k_z < k_z^{D}$ (c) $k_z = k_z^{D}$, and (d) $k_z > k_z^{D}$. Color representation of the orbitals is displayed on top of (a).}}
\label{band-NM}
\end{figure}

The first-principles calculations are performed using Quantum Espresso \cite{QUAN,Baro} with an 18$\times$18$\times$6 $\mathbf{k}$-grid. Norm-conserving PBE functionals are used in the calculations with an energy cut-off being 100 Ry. The SOC is included in all calculations. The Wannier functions obtained from the Wannier90 code \cite{Most} are used to compute the hopping constants for the effective Hamiltonian.
\begin{figure}[h]
\includegraphics [width=7.8cm] {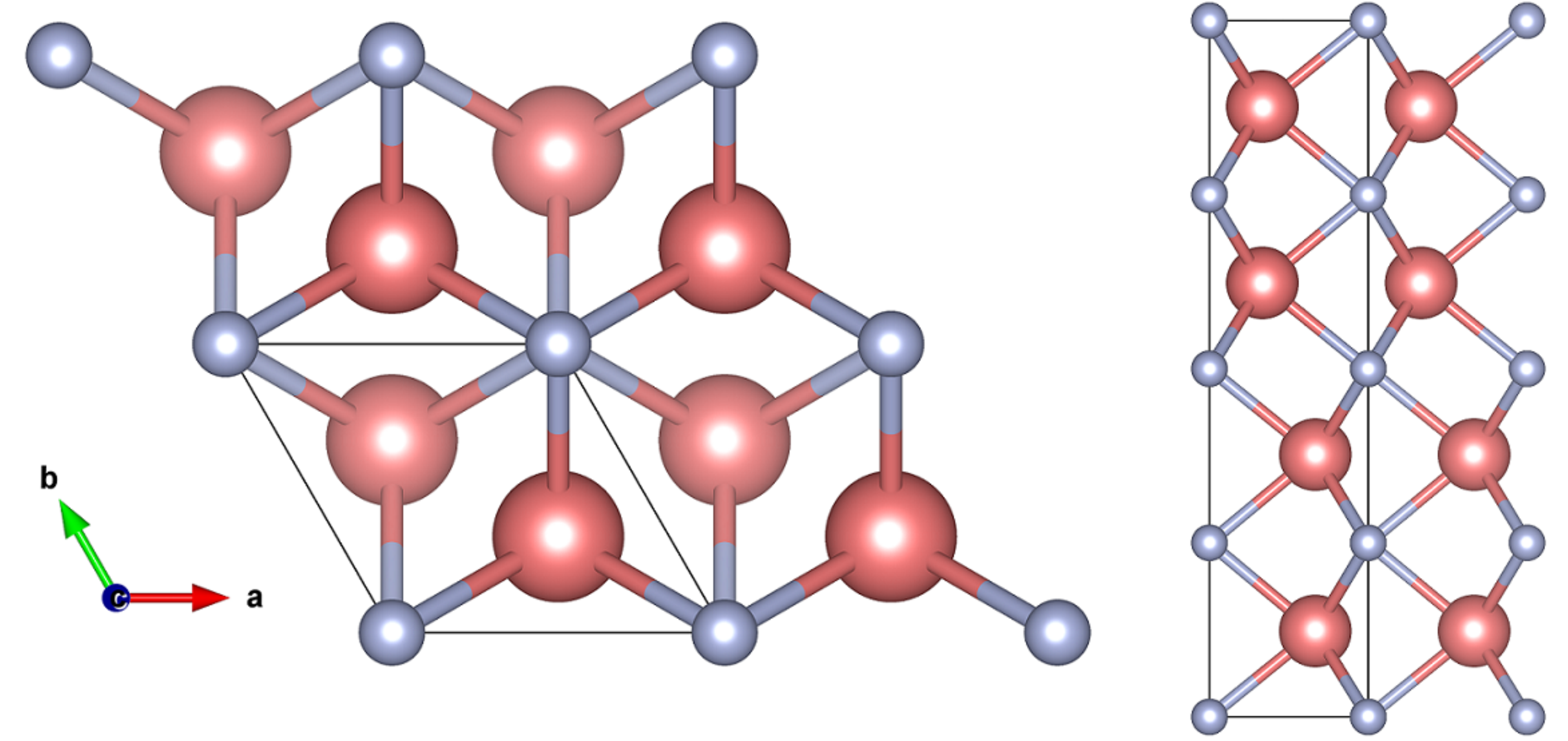}
\caption{\small{The crystal structure of the $\varepsilon$-TaN. Larger pink (smaller gray) spheres represent the Ta (N) atoms. Left and right panels display the top and side views, respectively. The lattice parameters are $a = b =$ 2.951 {\AA} and $c =$ 11.355 {\AA} from our structural relaxation.}}
\label{crystal}
\end{figure}
Figure \ref{crystal} displays the crystal structure of $\varepsilon$-TaN ($P6_{3}/mmc$, space group No. 194) with inversion and six-fold screw symmetries. The band structure calculated by the density functional theory is shown in Fig. \ref{band-NM}. 
It reveals that the Ta-$d$ orbitals are responsible for the states around the Fermi level, especially the $|J = \frac{5}{2}; J_z = \pm\frac{5}{2}\rangle$ and $|J = \frac{3}{2}; J_z = \pm\frac{3}{2}\rangle$ orbitals (abbreviated as $|\frac{5}{2}; \pm\frac{5}{2}\rangle$ and $|\frac{3}{2}; \pm\frac{3}{2}\rangle$ hereafter). The TNLs are marked by the arrows in Fig. \ref{band-NM}(a). It can be seen that there are two TNLs whose crossing points at $A$ are close in energy, and the crossing of their split bands on the $k_z$-axis results in a Dirac point [inset of Fig. \ref{band-NM}(a)], making $\varepsilon$-TaN a 3D Dirac system. Based on the topological classification reported in Ref. \cite{Yan1}, $\varepsilon$-TaN is a topological Dirac system (see Supplemental Materials \cite{supp} for details.).

In Ref. \cite{Yan1}, a four-band model is adopted and all relevant symmetry operations, e.g. crystalline and inversion, couple states within this four-band manifold. In $\varepsilon$-TaN, it is interesting and important to mention that there are eight (four doubly degenerate) bands around the Dirac point; four of them comprise of the two crossing bands and the other four are their ``folded bands'' due to the screw symmetry. Because of the special nonsymmorphic feature of space group 194, inversion symmetry, together with translation symmetry along the screw axis, will map a state originally in the crossing branch into a state in the folded branch, and vice versa \cite{Dres}. As a result, we need to consider all the eight bands for completeness. Important to mention, the extension from a four-band model to an eight-band model plays a critical role in the exotic topological properties that we discover in $\varepsilon$-TaN.

The electronic structure around the Dirac point reveals more interesting phenomena. As shown in Fig. \ref{band-NM}(a), the two Dirac bands along $\Gamma$-A show linear behavior, whereas they exhibit band dispersion higher than quadratic order along $k_y$ (as well as $k_x$, not shown), implying a high-order Dirac point. Figures \ref{band-NM}(b)-\ref{band-NM}(d) reveal the band inversion between $|\frac{5}{2}; \pm\frac{5}{2}\rangle$ and $|\frac{3}{2}; \pm\frac{3}{2}\rangle$ at $|k_z| < |k_z^{D}|$, where $k_z^D \simeq 0.348\cdot \frac{2\pi}{c} \simeq 0.192 ~\mathrm{\AA}^{-1}$ denotes the $k_z$ coordinate of the Dirac point. (The momentum will be expressed in unit of $2\pi/c$ hereafter unless otherwise mentioned.) Due to the ``gap closing'' at $k_z^D$, the band inversion is absent when $|k_z| > |k_z^{D}|$. This is consistent with the result that the plane with $k_z$ = 0 exhibits nontrivial $Z_2$ while that with $k_z = \pi/c$ is trivial, as is shown in the Supplemental Materials \cite{supp}. Careful inspection of Fig. \ref{band-NM}(c) on the way of band crossing manifests the cubic band dispersion, thus it is a cubic Dirac point following the convention \cite{Yan1}. This is different from what was found in Ref. \cite{Yan1}, there the cubic Dirac point is found at a high symmetry point like $A$ when only four-band manifold is considered. In other words, we argue that cubic Dirac points could be stabilized on the $k_z$-axis when it involves coupling of bands in a larger manifold. Thus, a different classification table is expected because of the more complicated coupling here. We would like to mention that such a cubic Dirac point also exists in $\mathrm{Na_3Bi}$ ($\sim$ 3 eV above the Fermi level), another binary compound with space group 194. Our work thus provides a way to search for new cubic Dirac semimetals.

When the time-reversal symmetry is broken, a Dirac point can be split into pairs of WPs. Here we apply a Zeeman field along the screw axis (or $k_z$ axis) to $\varepsilon$-TaN to study how the high-order Dirac point is split into the WPs. 
\begin{figure}[h]
\includegraphics [width=7.8cm] {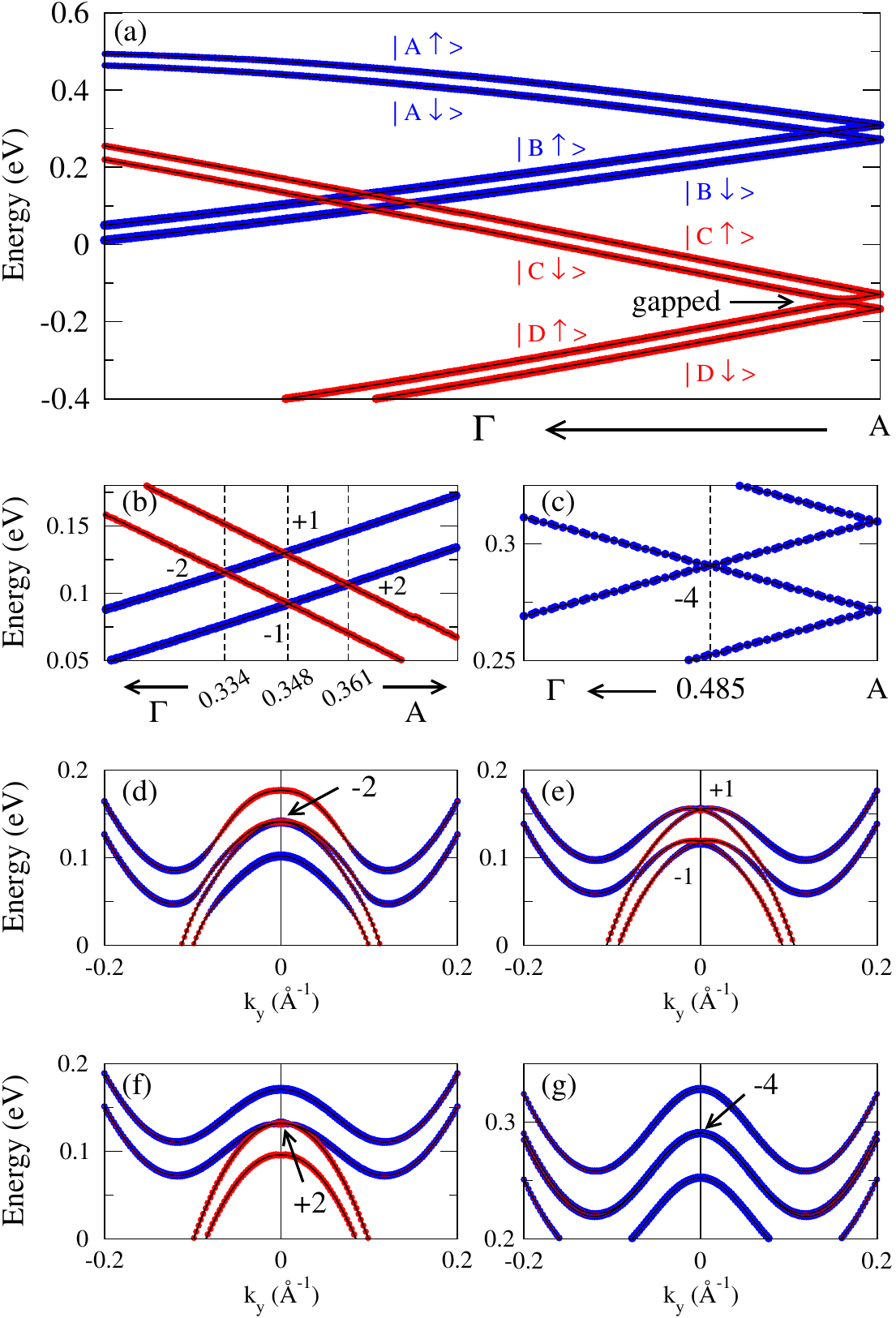}
\caption{\small{The band structures around the WPs of $\varepsilon$-TaN with a Zeeman field along the screw axis (a)-(c) along $k_z$ with different ranges in momentum and energy, (d) along $k_y$ with $k_z^{(-2)} = 0.334$, (e) along $k_y$ with $k_z^{(+1)}$ = 0.347, (f) along $k_y$ with $k_z^{(+2)} = 0.361$, and (g) along $k_y$ with $k_z^{(-4)} = 0.483$. The color representation of bands is the same as that in Fig. \ref{band-NM}. Labels of orbital in (a) are used in Table \ref{table:screw-eigen} to represent their screw eigenvalues.}}
\label{band-mag}
\end{figure}
Shown in Fig. \ref{band-mag} is the band splitting when a Zeeman field ($\frac{n_{\uparrow} - n_{\downarrow}}{n_{\uparrow} + n_{\downarrow}} = 0.01$) along the screw axis is applied to $\varepsilon$-TaN. The labeling of the eight bands is displayed in Fig. \ref{band-mag}(a). In addition to the Zeeman splitting of bands, there remain several WPs on the $k_z$-axis, indicating the topological phase transition into the Weyl state. The computed chiral charges are displayed in Figs. \ref{band-mag}(b)-(g). Because the WPs with $C =  +1$ and $C = -1$ occur at slightly different $k_z$, a small gap is seen in Fig. \ref{band-mag}(e) around the WP with $C = -1$. The band structure along $k_y$ with $k_z^{(-1)} = 0.348$ can be found in Supplemental Materials \cite{supp}, where the formation of a WP is manifest. The application of a Zeeman field perpendicular to the screw axis that breaks the screw symmetry, on the contrary, gives rise to ordinary Zeeman splitting of bands without showing any WPs. As a result, we focus on the effect of the Zeeman field along the screw axis in this work.

There are several ways of computing the chiral charge of a crossing point. In this work, we adopt three different methods to confirm the chiral charges, including the calculations of Wilson loop, Chern number, and the comparison of rotation (screw) eigenvalues of the conduction and valence bands \cite{Fang}. 
\begin{figure}[h]
\includegraphics [width=7.8cm] {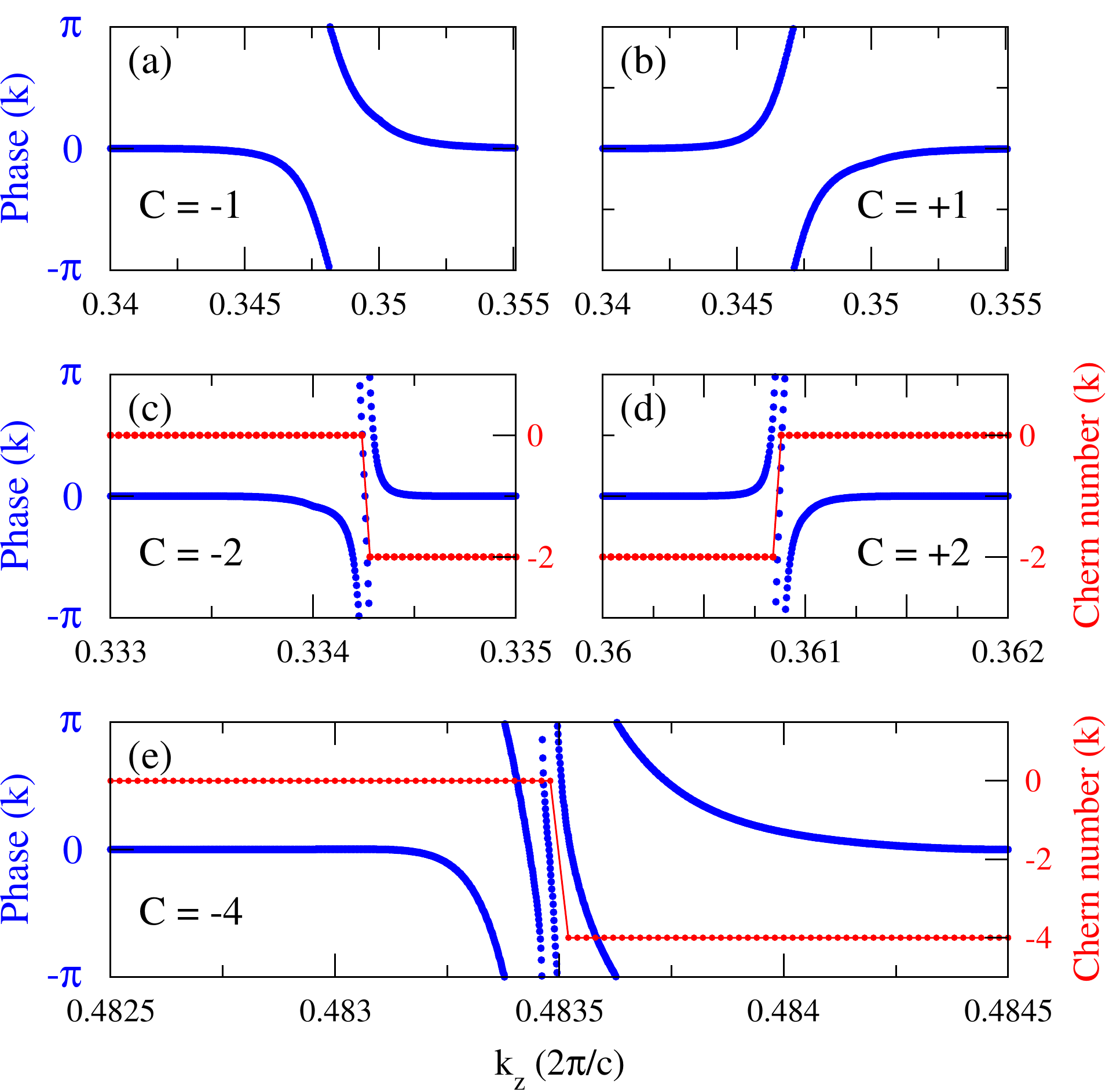}
\caption{\small{The chiral charges of the WPs in $\varepsilon$-TaN with a Zeeman field along the screw axis. (a) $C = -1$ at $k_z^D$. (b) $C = +1$ at $k_z^D$. (c) $C = -2$ at $k_z = 0.334$. (d) $C = +2$ at $k_z = 0.361$. (e) $C = -4$ at $k_z = 0.483$. The phases calculated from the Wilson loop are plotted in blue (left axis) and the Chern numbers are in red (right axis).}}
\label{chiral}
\end{figure}
As shown in Fig. \ref{chiral}, the chiral charges of these WPs can be found from the winding numbers in the Wilson loops (blue dots). Apparently, the four WPs split from the Dirac point exhibit $C = \pm 1, \pm 2$, respectively, which is consistent with the cubic DP.  More intriguing is that the WP formed due to the additional band crossing near $A$ point exhibits high $C = -4$ (There is another WP with $C = 4$ located at $-k_z$ so that the sum of chiral charges within the Brillouin zone is zero.). Note that the doubly degenerate quadruple WP we find is expected to have different Landau levels as compared to the reported highly degenerate WPs \cite{bradlyn2016,schroter2019,chang2017,Wu}, due to distinct band dispersion and hence the density of states. This in turn may lead to different transport properties such as anomalous Hall conductivities and magneto-transport. In Figs. \ref{chiral}(c)-\ref{chiral}(e), the Chern number is also computed to check the chiral charge. Apparently, the changes in Chern number across these WPs also indicate $C = \pm 2$ and $C = -4$, respectively.

Another method proposed by Fang {\it et. al.} \cite{Fang} is also adopted to further confirm the computed chiral charges of the WPs. It is stated that the ratio of rotation eigenvalues $\frac{u_c}{u_v} = \frac{e^{i\theta_1}}{e^{i\theta_2}} = e^{i\Delta\theta} = e^{i2\pi C/N}$ of the two crossing bands determines the chiral charge of that WP in a system with $N$-fold rotational symmetry. Due to the fact that $\theta$ is not uniquely determined in $e^{i\theta} = Z$, the chiral charge so obtained is actually $C$ mod $N$ \cite{supp}. Such scenario is later shown to be applicable to systems with screw symmetry \cite{Tsir}. The phases of the screw eigenvalues of the eight bands labeled in Fig. \ref{band-mag}(a) are listed in Table \ref{table:screw-eigen}. 
\begin{table}[h]
\begin{center}
\caption{\small{The calculated phases ($\theta = \frac{n\pi}{3}$), of the screw eigenvalues ($\lambda = e^{i\theta}$) of the eight bands labeled in Fig. \ref{band-mag}(a). Difference in $n$ between two crossed bands indicates the chiral charge of the WP.}}
\label{table:screw-eigen}
\begin{ruledtabular}
\begin{tabular}{ccccccccc}
band & $|A\uparrow\rangle$ & $|A\downarrow\rangle$ & $|B\uparrow\rangle$ & $|B\downarrow\rangle$ & $|C\uparrow\rangle$ & $|C\downarrow\rangle$ & $|D\uparrow\rangle$ & $|D\downarrow\rangle$ \\ \hline
   $n$    &  -2   &   5   &   1   &   2   &   0   &   3   &   3   &   0 \\
\end{tabular}
\end{ruledtabular}
\end{center}
\end{table}
Consistent chiral charges of all WPs can be obtained. Details of the model used to obtain these screw eigenvalues can be found in Supplemental Material \cite{supp}. In order to further confirm that WPs with $|C|=4$ are stable against symmetry-allowed perturbations, we study the distorted systems where the glide symmetry is broken but the protecting symmetry, the screw symmetry, is kept. As shown in the Supplemental Material \cite{supp}, the results of Wilson-loop calculations for two distorted systems indicate that the chiral charges are unchanged under the distortions as expected. We also confirm that the calculated $C = -4$ is contributed from a single WP rather than several ones with total $C$ adding up to $-4$ \cite{supp}. Although the chiral charges of the WPs are confirmed, the topological edge states, on the other hand, are difficult to identify because all these WPs are too close to each other in both energy and momentum space. Also, the edge states around the WPs are buried in the bulk bands because of the cubic band dispersion. Despite this, the nontrivial WPs are still expected to contribute to bulk properties such as (magneto-)transport. As mentioned, $\varepsilon$-NbN reveals the same topological behaviors and the results are shown in Supplementary Materials \cite{supp}.

\section{Discussions}
Despite that a WP exhibiting $|C| = 4$ is surprising, its presence could be understood from the following argument. Recall that the TNLs pass through $A$ and become doubly degenerate under the application of a Zeeman field along the screw axis. As pointed out in Refs. \cite{Lian,Yan2} on materials belonging to space group 194, the calculated winding number, defined as $\gamma/\pi$ with $\gamma$ being the Berry phase, of each doubly degenerate TNLs is $\pm1$. At $A$ where three TNLs (three rotationally equivalent lines along $A$-$L$) spread out, it is expected that the increase in the phase ($\theta$) of the screw eigenvalues upon band folding would be $\pm\pi$ since each TNL, playing a similar role as a single chiral charge, contributes $\pm\frac{\pi}{3}$. Indeed, the differences of phases as shown in Table \ref{table:screw-eigen} indicate that the four crossings at $A$ act effectively as topological points with $|C| = 3$. That is, the TNLs put additional constraint on the phase winding of these bands. This explains why the phase of $|A\downarrow\rangle$ can be $\frac{5\pi}{3}$ in TaN because it acquires $\pi$ when folded from $|B\downarrow\rangle$. On the other hand, the TNLs contribute $-\pi$ to $|A\downarrow\rangle$ in $\mathrm{Na_3Bi}$ and hence $C = 2$ is obtained (see Supplemental Materials \cite{supp}). Having the same space group and band representations, our results indicate that symmetry does not fully account for the chiral charge of this WP. Rather, the chiral charge is determined by the coupling of bands in the eight-band manifold via the relative phase winding, which seems to be correlated to the parity change from $\Gamma$ to $A$ as discussed below.

In the eight-band manifold under consideration, \{$|A\rangle, |C\rangle$\} and \{$|B\rangle, |D\rangle$\} [see Fig. \ref{band-mag}(a)] form two four-fold degenerate bands, denoted respectively by $|U\rangle$ and $|L\rangle$, without SOC and Zeeman field. $|U\rangle$ and $|L\rangle$ are degenerate at $A$ with mixing parity of $+$ and $-$. At $\Gamma$, however, there are two possible situations. First, $|U\rangle$ and $|L\rangle$ show opposite parity, which is the case of $\mathrm{Na_3Bi}$. In this case, parity change is not required when going from $\Gamma$ to $A$ to reach the mixed parity state at $A$. The other case is that $|U\rangle$ and $|L\rangle$ have the same parity at $\Gamma$ as in TaN and NbN. Under these circumstances, parity mixing of both $|U\rangle$ and $|L\rangle$ must be so strong as to reach the mixed parity of $+$ and $-$ at $A$. As the band inversion in topological insulators gives rise to nonzero winding number of the split-off Wannier centers \cite{Fu2,Yu}, it is intuitive to speculate that the parity mixing imposes a relative winding of $2\pi$ on the screw eigenvalues, which is expected to contribute $\pm 6$ to the chiral charge in a six-fold screw symmetric system. The interplay between screw eigenvalues and parity may need further theoretical studies to clarify.

The coupling of the eight bands has another consequence on the chiral charges actually. In addition to the possible correlation between the parity and screw eigenvalues, we further find that the chiral charge of this high-order WP can only be either $-4$ or $+2$ ($+4$ or $-2$ at -$k$) in systems belonging to space group 194. It is noted that the WPs are formed at crossings of bands with either $J_z = \pm\frac{5}{2}$ (TaN and NbN) or $J_z = \pm\frac{1}{2}$ ($\mathrm{Na_3Bi}$). Then, the difference in screw eigenvalue of the two Zeeman-split bands (e.g. $|B\uparrow\rangle$ and $|B\downarrow\rangle$) is always $\pm \frac{\pi}{3}$ (see Supplemental Materials for details \cite{supp}). Considering the constraint imposed by the TNLs mentioned previously, the difference in the screw eigenvalues will be either $\pm\frac{4\pi}{3}$ or $\pm\frac{2\pi}{3}$, resulting in $C =-4 (+4)$ as in TaN and NbN or $C = 2 (-2)$ as in $\mathrm{Na_3Bi}$, because $|A\downarrow\rangle$ acquires a phase of $\pi$ or $-\pi$ when folded from $|B\downarrow\rangle$ at $A$. Since the acquisitions of $\pi$ and $-\pi$ in phase are both symmetry-respected, symmetry cannot fully account for the chiral charge of this high-order WP. This analysis provides another support for the eight-band model since the screw eigenvalues of the ``parent'' and ``folded'' branches are connected due to the TNLs.

Screw symmetry is the main ingredient of the materials exhibiting high-order WP (such as $|C|$ = 4 in this work). The presence of screw symmetry protects the band folding at the BZ boundary along the screw axis and thus results in high band degeneracy (such as $A$-point in TaN with four-fold degeneracy). This in turn leads to the finite slope of the band dispersion around the folding point (or crossing point when extending over the zone boundary). The importance of having finite slope of bands is that the band characters interchange upon passing the crossing point at the zone boundary, which is related to the vanishing structure factor at $A$ \cite{Dres}. Suppose two sets of such folded bands are close in energy, which is naturally satisfied if the two sets of bands are split by SOC, the minimum model has to cover all the coupled bands, i.e. an eight-band manifold in this work, so as to satisfy the closeness of the inversion operation. The complicated couplings in a larger manifold of bands can give rise to high-order WPs. Therefore, one can follow this direction to search for other materials showing high-order WPs, e.g. BaTaS$_3$ \cite{Lian} in which we find three sets of SOC-split valence bands at $A$.

In conclusion, we have demonstrated that $\varepsilon$-TaN is a 3D topological material showing cubic Dirac point. Upon the application of a Zeeman field along the screw axis, this Dirac point splits into WPs with chiral charges $C = \pm 1, \pm2$, respectively. An extra band crossing gives rise to a high-order WP carrying $|C| = 4$.
Importantly, an eight-band model is required to describe the Dirac point as well as the split WPs. Moreover, symmetry alone cannot fully determine the chiral charge of this high-order WP. Instead, it is determined by symmetry combined with the band-mixing within this eight-band manifold.

\section{Acknowledgements}
The authors thank Xiangang Wan, Ching-Kai Chiu, and Tay-Rong Chang for fruitful discussions. P.-J.C. and T.-K.L acknowledge National Center for High-Performance Computing (NCHC). This work is supported by Ministry of Science and Technology and Taiwan Consortium of Emergent Crystalline Materials (TCECM) (Grant No. MOST 106-2119-M-001-028) and Academia Sinica.

\section{Author Constributions}
P.J.C and W.J.L. performed the numerical calculations, analysed the data, and wrote the paper. T.K.L. supervised the project. P.-J.C. and W.-J.L. contributed equally to this work.

{}

\end {document}